\documentclass[prd,preprintnumbers,nofootinbib,reprint,twocolumn]{revtex4-1}
\usepackage{amsmath,amssymb,bm,epsfig,color,graphicx}
\usepackage[mathscr]{eucal}
\usepackage{slashed}
\usepackage{cancel}
\usepackage{yhmath}
\usepackage[colorlinks=true,linkcolor=blue,citecolor=blue,urlcolor=blue]{hyperref}
\usepackage[caption=false]{subfig}
\usepackage{hyperref}
\usepackage{amsmath}
\usepackage{slashed}
\usepackage{comment}
\usepackage{cancel}
\usepackage{multirow}

\newcommand{\ptoKv}{p \to K^+ \bar{\nu}}

\begin{document}

\title{Proton Lifetime in Minimal Supersymmetric SU(5) with Gauge Mediation}
\author{Jason L. Evans$^{1,2}$ and Yoshihiro Shigekami$^{3}$\\*[10pt]
$^1${\it \normalsize Tsung-Dao Lee Institute, Shanghai Jiao Tong University,\\
No. 1 Lisuo Road, Pudong New Area, Shanghai 201210, China} \\*[3pt]
$^2${\it \normalsize School of Physics and Astronomy, Shanghai Jiao Tong University,\\
800 Congchuan Road, Shanghai 200240, China} \\*[3pt]
$^3${\it \normalsize Institute of Particle and Nuclear Physics, Henan Normal University,\\
46 East of Construction Road, Xinxiang, Henan, 453007, China}\\*[5pt]
}

\begin{abstract}
In this paper, we discuss the predicted proton lifetimes in minimal supersymmetric (SUSY) $SU(5)$ grand unified theory (GUT) with gauge mediated supersymmetry breaking (GMSB). 
We focus on the case of $\mathbf{5} + \mathbf{\bar{5}}$ messengers and determine the low-scale mass spectrum of the scalar particles and gauginos using the renormalization group equations. 
With the obtained mass spectrum, we calculate the dominant proton decay mode for SUSY $SU(5)$ GUT, $\ptoKv$. 
In our setup, we assume the messenger scale to be $\mathcal{O}(10^3)$ TeV in order to obtain a proper Higgs mass in GMSB scenario. 
For this messenger scale, we find the proton lifetime is consistent with current experimental limits and can be tested by future proton decay experiments. 
\end{abstract}

\maketitle

\section{Introduction}
\label{sec:intro}

It is well-known that supersymmetry (SUSY) is one of the most promising extensions of the standard model (SM). 
It predicts a natural electroweak scale with a 125 GeV Higgs mass and predicts a neutral and stable particle which is a good candidate for dark matter. 
Additionally, supersymmetric models generally predict precise gauge coupling unification unlike SM. 
It is hard to believe that this extremely precise unification, which is dependent on the experimentally measured gauge couplings, is just due to coincidence. 
This is strongly suggestive of a grand unified theory (GUT)~\cite{Pati:1973rp,Pati:1974yy,Georgi:1974sy,Georgi:1974yf,Georgi:1974my,Fritzsch:1974nn} where the three SM gauge groups are unified. 
Furthermore, in unified theories, the matter fields are embedded into larger multiplets, e.g., for $SU(5)$ GUT models an entire generation of the standard model can be embedded into $\mathbf{10} + \mathbf{\bar{5}}$. 
This embedding leads to correlations among the different charges and is an explanation for charge quantization. 

Although these generic feature of SUSY GUT models are compelling, there are challenges in developing a viable model. 
One particularly important concern is SUSY breaking. 
When SUSY is broken, generically, there are lots of new couplings and phases introduced. 
These couplings can generate low-scale flavor and CP violation. 
If the SUSY soft masses are around the weak scale~\cite{CMS:2021beq,ATLAS:2021jyv,CMS:2023xlp}, the predicted flavor and CP violation exceed the experimental bounds. 
Even the soft masses are larger and consistent with current collider experiments, it is still difficult to avoid the constraints coming from experiments looking for flavor and CP violation. 
This is known as a SUSY flavor and CP problem. 
Although this problem can be solved by taking the soft masses to be well beyond current experimental bounds, the constraints are so severe that this introduces a little hierarchy problem. 

To solve the SUSY flavor and CP problem in the context of minimal SUSY $SU(5)$ GUT models, we utilize gauge mediated SUSY breaking (GMSB)~\cite{Dine:1981za,Dimopoulos:1981au,Dine:1981gu,Nappi:1982hm,Alvarez-Gaume:1981abe,Dine:1993yw,Dine:1994vc,Dine:1995ag}\footnote{For review, see Refs.~\cite{Giudice:1998bp,Kitano:2010fa}.}. 
In the case of GMSB, we obtain a flavor universal mass spectrum for sfermions, which helps alleviate the dangerous of flavor changing processes\footnote{Gauge mediation also provides a possible means to solve the CP problem dynamically \cite{Evans:2010ru}. 
However, we do not pursue this here.}. 

Since the low-scale masses of the minimal SUSY SM (MSSM) particles in GMSB and gravity mediated SUSY breaking tend to be of the same order to explain the Higgs mass\footnote{Both require a stop mass of order 10~{\rm TeV}~\cite{Okada:1990vk,Ellis:1990nz,Haber:1990aw,Okada:1990gg,Ellis:1991zd,Ibe:2012qu,Asano:2016oik,Bhattacharyya:2018inr} aligning the scales of the soft masses.}, the proton decay prediction for the decay mode $\ptoKv$, which is partially mediated by the supersymmetry breaking, should generically be of the same order in the two models. 
Thus, we expect a proton lifetime of order $10^{34}$ years in GMSB model, which is within reach of coming experiments~\cite{Bhattiprolu:2022xhm}. 

In this paper, we calculate the proton lifetime and explore the viable parameter space of GMSB when both the Higgs mass, 125 GeV, and proton lifetime, for $\ptoKv$ process, are considered. 
Moreover, we also discuss the messenger mass scale dependency of the predicted $\ptoKv$ decay mode. 
We also examine detection of this mode at future proton decay experiments: Hyper-Kamiokande~\cite{Hyper-Kamiokande:2018ofw}, Deep Underground Neutrino Experiment (DUNE)~\cite{DUNE:2015lol} and Jiangmen Underground Neutrino Observatory (JUNO)~\cite{JUNO:2015zny}. 

The paper is organizing as follows. 
In section~\ref{sec:model}, we summarize visible sector and SUSY breaking sector of our model. 
Section~\ref{sec:lifetime} gives relevant information about the calculation of the proton lifetime for $\ptoKv$ process. 
We give our numerical results in section~\ref{sec:results}, and then, summarize in section~\ref{sec:summary}. 
In appendix~\ref{sec:calcinfo}, the auxiliary information for the calculation of proton lifetime can be found.

\section{Model description}
\label{sec:model}

In this paper, we focus on the minimal SUSY $SU(5)$ model with GMSB for a pair of $\mathbf{5} + \mathbf{\bar{5}}$ messengers. 
Here, we will explain our model.

\subsection{The Grand Unified Theory}
\label{sec:visible}

We consider the following $SU(5)$ invariant superpotential\footnote{In our study, we omit other higher dimensional operators assuming that they should contribute at a subleading order. 
In order for the neglected higher dimensional operators to be important, they would need to dominate the tree-level contribution.}:
\begin{align}
W_{\rm vis} &= \mu_{\Sigma} {\rm Tr} \Sigma^2 + \frac{\lambda'}{6} {\rm Tr} \Sigma^3 + \mu_H \overline{H} H + \lambda \overline{H} \Sigma H \nonumber \\[0.5ex]
&\hspace{1.2em} + (h_{\mathbf{10}})_{i j} \mathbf{10}_i \mathbf{10}_j H + (h_{\mathbf{\bar{5}}})_{i j} \mathbf{10}_i \mathbf{\bar{5}}_j \overline{H} \, ,
\label{eq:Wvisible}
\end{align}
where $H$ and $\overline{H}$ are fundamental and anti-fundamental superfields containing the SM Higgs boson, and $\Sigma$ is an adjoint superfield whose vacuum expectation value (VEV) breaks $SU(5)$ down to SM gauge symmetries $SU(3)_C \times SU(2)_L \times U(1)_Y$. 
This breaking corresponds to $\langle \Sigma \rangle = V \cdot {\rm diag} (2, 2, 2, -3, -3)$, where $V = 4 \mu_{\Sigma} / \lambda'$. 
The $\mathbf{10}_i$ and $\mathbf{\bar{5}}_i$ contain the SM fermions with generational index $i = 1, 2, 3$, and $(h_{\mathbf{10}})_{i j}$ and $(h_{\mathbf{5}})_{i j}$ are the Yukawa couplings for SM fermions. 

The masses of the GUT scale field after $SU(5)$ is broken are labeled as $M_X$, $M_{H_C}$ and $M_{\Sigma}$ and correspond to the GUT theory gauge bosons, color triplet Higgs boson and color octet (and weak triplet) of $\Sigma$, respectively. 
They are found to be
\begin{align}
M_X = 5 g_5 V \, , ~~ M_{H_C} = 5 \lambda V \, , ~~ M_{\Sigma} = \frac{5}{2} \lambda' V \, .
\label{eq:HeavyMasses}
\end{align}

In order to get a viable unified theory, the gauge couplings must unify. 
The masses in Eq.~\eqref{eq:HeavyMasses} represent incompletely multiplets of the unified theory and so lead to deviations in the running of the gauge couplings. 
In order for the SM measured values of the guage coupling to be consistent with unification, the gauge couplings must satisfy the following conditions~\cite{Hisano:1993zu,Tobe:2003yj,Evans:2015bxa}
\begin{align}
\frac{1}{g_1^2 (Q)} &= \frac{1}{g_5^2 (Q)} + \frac{1}{8 \pi^2} \left[ \frac{2}{5} \ln \frac{Q}{M_{H_C}} - 10 \ln \frac{Q}{M_X} \right] \nonumber \\[0.3ex]
&\hspace{1.2em} - \epsilon \, , \label{eq:g1match} \\[0.5ex]
\frac{1}{g_2^2 (Q)} &= \frac{1}{g_5^2 (Q)} + \frac{1}{8 \pi^2} \left[ 2 \ln \frac{Q}{M_{\Sigma}} - 6 \ln \frac{Q}{M_X} \right] \nonumber \\[0.3ex]
&\hspace{1.2em} - 3 \epsilon \, , \label{eq:g2match} \\[0.5ex]
\frac{1}{g_3^2 (Q)} &= \frac{1}{g_5^2 (Q)} + \frac{1}{8 \pi^2} \left[ \ln \frac{Q}{M_{H_C}} + 3 \ln \frac{Q}{M_{\Sigma}} - 4 \ln \frac{Q}{M_X} \right] \nonumber \\[0.3ex]
&\hspace{1.2em} + 2 \epsilon \, , \label{eq:g3match}
\end{align}
where $g_i$ are the SM gauge couplings and the $U(1)_Y$ gauge coupling has the GUT normalization $g_1^2 = (5/3) g_Y^2$. 
The $\epsilon$ ($ \equiv 8 d V / M_P$) in Eqs.~\eqref{eq:g1match}-\eqref{eq:g3match} is a contribution from the following important Planck suppressed operator~\cite{Ellis:1979fg,Panagiotakopoulos:1984wf,Nath:1996qs,Nath:1996ft,Bajc:2002pg},
\begin{align}
W_{\rm eff}^{\Delta g} = \frac{d}{M_P} {\rm Tr} \left[ \Sigma \mathcal{W} \mathcal{W} \right] \, , \label{eq:Weff}
\end{align}
where $M_P$ is the Planck mass and the $SU(5)$ field strength superfield is defined as $\mathcal{W} = \mathcal{W}^A T^A$. 
It is noteworthy that the size of $\epsilon$ is comparable to the one-loop threshold corrections due to the fact that $V / M_P \simeq 10^{-2}$, and hence, must be included in any precision numerical analysis. 
This $\epsilon$ contribution in Eqs.~\eqref{eq:g1match}-\eqref{eq:g3match} is important since it allows the Higgs color triplet mass to be a free parameter,
\begin{align}
\epsilon &= \frac{1}{12} \left( \frac{1}{g_1^2 (Q)} - \frac{3}{g_2^2 (Q)} + \frac{2}{g_3^2 (Q)} \right) - \frac{1}{40 \pi^2} \ln \frac{Q}{M_{H_C}} \nonumber \\[0.5ex]
&\to \frac{1}{6 g_3^2 (M_{\rm GUT})} - \frac{1}{6 g_1^2 (M_{\rm GUT})} - \frac{1}{40 \pi^2} \ln \frac{M_{\rm GUT}}{M_{H_C}} \, ,
\end{align}
where the second line is obtained by setting $Q = M_{\rm GUT}$, with $M_{GUT}$ defined as the solution to $g_1 (M_{\rm GUT}) = g_2 (M_{\rm GUT})$. 
Note that in addition to $M_{H_C}$ beiging completely determined when $\epsilon = 0$, the three conditions in Eqs.~\eqref{eq:g1match}-\eqref{eq:g3match} must be solved by four parameters, $(V, \lambda, \lambda', g_5)$, leaving only one free parameter. 
With only one free parameter it is difficult to find viable models.

\subsection{Messenger sector}
\label{sec:messenger}

Now, we consider the messenger sector of GMSB. 
We will focus on the case of $\mathbf{5} + \mathbf{\bar{5}}$ messenger multiplets~\cite{Ibe:2012qu,Asano:2016oik,Bhattacharyya:2018inr}:
\begin{align}
W_{\rm mess} &= (M_L + k_L Z) \Psi_L \Psi_{\bar{L}} + (M_D + k_D Z) \Psi_D \Psi_{\bar{D}} \nonumber \\
&\hspace{1.2em} - \xi_Z Z\, ,
\label{eq:WMes}
\end{align}
where $\Psi_L$ and $\Psi_D$ are the $SU(2)_L$ doublet and $SU(3)_C$ triplet messengers, and $Z$ is a singlet superfield which breaks SUSY\footnote{Although GUT theories dictate that $M_L$ and $M_D$ arise from the same mass term, $M \Psi \bar{\Psi} = M (\Psi_D \Psi_{\bar{D}} + \Psi_L \Psi_{\bar{L}})$, we can achieve $M_L \neq M_D$ by introducing additional terms of the form $\lambda_{\Psi} \bar{\Psi} \Sigma \Psi$. 
Once $\Sigma$ gets a VEV, the masses are split, $M_L = M - 3 \lambda_{\Psi} V$ and $M_D = M + 2 \lambda_{\Psi} V$. 
By taking $\lambda_\Psi\ll 1$, independent masses with $M_L\sim M_D$ can be realized at any scale.}. 
When integrated out, these messengers generate the following gaugino and scalar masses 
\begin{align}
M_1 &\simeq \frac{g_1^2}{16 \pi^2} \left( \frac{2}{5} \Lambda_D + \frac{3}{5} \Lambda_L \right) \, , \label{eq:M1GMSB} \\
M_2 &\simeq \frac{g_2^2}{16 \pi^2} \Lambda_L \, , \quad M_3 \simeq \frac{g_3^2}{16 \pi^2} \Lambda_D \, , \label{eq:M2M3GMSB} \\[0.5ex]
m_{\varphi_i}^2 &\simeq \frac{2}{(16 \pi^2)^2} \left[ C_3 (i) g_3^4 \Lambda_D^2 + C_2 (i) g_2^4 \Lambda_L^2 + \frac{3}{5} Y_i^2 g_1^4 \Lambda_Y^2 \right] \, , \label{eq:mphiGMSB}
\end{align}
where $C_{2, 3} (i)$ are the quadratic Casimir invariants of the field $i$ for $SU(2)_L$ and $SU(3)_C$, respectively, $\Lambda_{L, D} \equiv k_{L, D} \langle F_Z \rangle / M_{L, D}$ with $\langle F_Z \rangle = \xi_Z$, and $\Lambda_Y^2 = (2/5) \Lambda_D^2 + (3/5) \Lambda_L^2$. 
We assume that the messenger masses are of the same order, i.e. $M_{\rm mess} \sim M_L \sim M_D$\footnote{These masses are corrected by the scalar vev of the supersymmetry breaking field, $\langle \phi_Z \rangle$, $M_{L, D} + k_{L, D} \langle \phi_Z \rangle$. 
However, the scalar vev of $Z$ is expected to be of order the SUSY breaking scale. 
In our calculation, we take $M_{\rm mess} = \mathcal{O} (10^3)$ TeV, while our SUSY breaking scale is $\mathcal{O} (10)$ TeV. Thus, our assumption of $M_{L, D} \gg k_{L, D} \langle \phi_Z \rangle$ is justified.}. 
Note that for the other SUSY breaking parameters, including $A$-terms, there are contributions from the messengers, but they are much smaller than the gaugino masses and so are taken to be zero in our numerical calculation. 

As studied in Refs.~\cite{Dvali:1996cu,Dine:1997qj,Langacker:1999hs,Hall:2002up,Roy:2007nz,Murayama:2007ge,Giudice:2007ca,Liu:2008pa,Csaki:2008sr,DeSimone:2011va}, it is known that in GMSB scenario it is difficulty to obtain an appropriately sized $\mu$ and $B_\mu$ parameters, the so called $\mu$-$B_\mu$ problem. 
The difficulty with these parameters arises from the fact that they come from different mechanisms. 
$\mu$ is a Higgs bilinear terms in the superpotential and $B_\mu$ is a Higgs bilinear in the scalar potential:
\begin{align}
W \supset \mu H_1 H_2 \, , \quad V_{pot} \supset B_\mu H_1 H_2 + {\rm h.c.} \, ,
\end{align}
where $H_{1, 2}$ are two Higgs doublet fields. 
In order to arrive at a viable model, this problem must be solved. 
However, in the context of GUT theories with low messenger scales, this problem can be easily solved. 
When combining GMSB with grand unification there is another problem with $\mu$ and $B_\mu$ associated with the GUT matching conditions found in \cite{Ellis:2017djk}. 
First, the $\mu_H$ must be tuned against a GUT scale contribution $\lambda V$ to get a weak scale Higgs doublet. 
This tuning gives us the freedom to choose $\mu$ as we wish. 
$B_\mu$, on the other hand, gets a gravity mediated contribution which is proportional to the gravitino mass, which is small for low scale messenger case. 
However, the gravity mediated contribution to $B_\mu$ is also enhanced by $\mu_H$ which must be of order the GUT scale. 
Generically, this gives $B_\mu\sim m_{3/2}M_{GUT}$, which is too large. 
This means we can use the same techniques used in gravity mediation found in \cite{Ellis:2017djk} to realize a viable $B_\mu$ parameter. 
Although this requires us to tune $B_\mu$, just like what is done in gravity mediated models, the required tuning is significantly less. 
Thus, GMSB models of $SU(5)$ have some advantages over gravity mediation as well as GMSB without grand unification.

\section{Proton lifetime}
\label{sec:lifetime}

With the above superpotential for the visible sector, the proton will decay mediated by the dimension-5 operators~\cite{Sakai:1981pk,Weinberg:1981wj}
\begin{align}
W_{p-{\rm decay}} &= \frac{1}{2} C_{5L}^{ijkl} \epsilon_{abc} \left( Q_i^a \cdot Q_j^b \right) \left( Q_k^c \cdot L_l \right)\nonumber \\[0.3ex]
&\hspace{2.4em}+ C_{5R}^{ijkl} \epsilon_{abc} \left( \bar{u}_{i a} \bar{e}_j \bar{u}_{k b} \bar{u}_{l c} \right) \, , \label{eq:Wpdec}
\end{align}
where coefficients $C_{5L, 5R}^{ijkl}$ are
\begin{align}
C_{5L}^{ijkl} & = \frac{\sqrt{8}}{M_{H_C}} h_{\mathbf{10}, i} e^{i \phi_i} \delta^{ij} V_{kl}^* h_{\mathbf{\bar{5}}, l} \, , \label{eq:C5Lijkl} \\[0.5ex]
C_{5R}^{ijkl} & = \frac{\sqrt{8}}{M_{H_C}} h_{\mathbf{10}, i} V_{ij} V_{kl}^* h_{\mathbf{\bar{5}}, l} e^{- i \phi_k} \, , \label{eq:C5Rijkl}
\end{align}
with $V_{ij}$ being the Cabibbo-Kobayashi-Maskawa (CKM) matrix element and $\phi_i$ being the additional GUT phases\footnote{These phases are additional degrees of freedom in the Yukawa couplings arising in $SU(5)$ grand unified theories. 
We start with $12 + 18$ parameters in $(h_{\mathbf{10}})_{ij}$ and $(h_{\mathbf{\bar{5}}})_{ij}$. 
Through allowed field redefinitions, which are $U(3) \otimes U(3)$ transformation, it is seen that the actual degrees of freedom are $(12 + 18) - 9 \times 2 = 12$. 
These 12 parameters include 6 quark masses, 4 parameters for the CKM matrix, and 2 additional phases.}, which satisfy the condition $\phi_1 + \phi_2 + \phi_3 = 0$. 
Note that $h_{\mathbf{10}, i}$ and $h_{\mathbf{\bar{5}}, i}$ are eigenvalues of $\left( h_{\mathbf{10}} \right)_{ij}$ and $\left( h_{\mathbf{\bar{5}}} \right)_{ij}$, respectively, and in our analysis, we take the following basis for $h_{\mathbf{10}}$ and $h_{\mathbf{\bar{5}}}$ as~\cite{Hisano:1992jj}
\begin{align}
\left( h_{\mathbf{10}} \right)_{ij} = e^{i \phi_i} \delta^{ij} h_{\mathbf{10}, i} \, , \quad \left( h_{\mathbf{\bar{5}}} \right)_{ij} = V_{ij}^* h_{\mathbf{\bar{5}}, j} \, .
\end{align}
To calculate the proton lifetime, we need to run these coefficients down to the QCD scale using appropriate renormalization group equations (RGEs). 
The resulting decay width of $\ptoKv_i$ can then be calculated giving
\begin{align}
\Gamma (\ptoKv_i) = \frac{m_p}{32 \pi} \left( 1 - \frac{m_K^2}{m_p^2} \right)^2 \left| \mathcal{A} (\ptoKv_i) \right|^2 \, , \label{eq:Gamp2Kvi}
\end{align}
where $m_p$ and $m_K$ are masses of the proton and the Kaon, and the corresponding amplitude $\mathcal{A} (\ptoKv_i)$ is discussed in appendix~\ref{sec:calcinfo}. 
In our numerical calculation, we sum all decay modes ($i = 1, 2, 3$) to obtain the lifetime of $\ptoKv$,
\begin{align}
\tau (\ptoKv) = \left( \sum_{i = 1}^3 \Gamma (\ptoKv_i) \right)^{-1} \, .
\label{eq:tauptoKv}
\end{align}
More details on the calculation for the proton lifetime can be found in Ref.~\cite{Ellis:2015rya}.

\section{Numerical results}
\label{sec:results}

In this section, we summarize our numerical results for the proton lifetime, $\ptoKv$, in the GMSB scenario. 
We assume that at the messenger scale, the gaugino and scalar masses are given by Eqs.~\eqref{eq:M1GMSB}-\eqref{eq:mphiGMSB}, and $A$-terms and $B$ parameter are subleading and so are set to zero. 
The RGEs then evolve the parameters to the low-scale\footnote{We use the SSARD code for this calculation~\cite{SSARDcode}.}. 
The $\mu$ and $B_\mu$ parameters are determined by the electroweak symmetry breaking (EWSB) conditions leaving eight free parameters\footnote{$k_{D, L}$ and $M_{D, L}$ at the messenger scale can be obtained by assuming $k_D = k_L (\equiv k)$, $M_D=M+2\lambda_\Psi V$, and $M_L=M-3\lambda_\Psi V$ at the GUT scale and renormalization group running the parameters to the messenger scale. Thus, $k$ and $M$ together with $\xi_Z$ and $\lambda_\Psi$ can be taken as the free parameters of the model. 
However, in our current study, we have chosen $\Lambda_{D, L}$ and $M_{\rm mess}$ to be free parameters of the model, since they are more directly related to proton decay. 
In principle, it is possible to reproduce any values of $(\Lambda_D, \Lambda_L, M_{\rm mess})$ from a suitable choice of $(k, M, \xi_Z,\lambda_\Psi)$.}:
\begin{align}
\tan \beta \, , \quad \Lambda_D \, , \quad \Lambda_L \, , \quad M_{\rm mess} \, , \quad \lambda \, , \quad \lambda' \, , \quad \phi_{2, 3} \, ,
\end{align}
where we have chosen $\phi_{2, 3}$ to be free parameters, and $\phi_1=- (\phi_2 + \phi_3)$. 
Note that $\lambda$ and $\lambda'$ play a role in determining $M_{H_C}$ and $M_{\Sigma}$ (see, Eq.~\eqref{eq:HeavyMasses}). 
Here, we set these to be $\lambda = 0.6$ and $\lambda' = 10^{-3}$. 
In previous studies of GMSB scenario~\cite{Okada:1990vk,Ellis:1990nz,Haber:1990aw,Okada:1990gg,Ellis:1991zd,Ibe:2012qu,Asano:2016oik,Bhattacharyya:2018inr}, it was shown that in order to reproduce the correct mass for the SM Higgs boson $\Lambda_D = \mathcal{O}(10^3)$ TeV. 
Therefore, we set $M_{\rm mess} = 1500$ TeV. 
Note, the proton lifetime is only mildly dependent on the choice of $M_{\rm mess}$. 
$\Lambda_{L, D}$ are varied to obtain a Higgs mass of 125 GeV. 
For the Higgs mass calculation, we use \texttt{FeynHiggs 2.18.1}~\cite{Heinemeyer:1998yj,Heinemeyer:1998np,Degrassi:2002fi,Frank:2006yh,Hahn:2013ria,Bahl:2016brp,Bahl:2017aev,Bahl:2018qog}\footnote{\href{http://www.feynhiggs.de/}{http://www.feynhiggs.de/}}. 
$\tan \beta$, the ratio of VEVs of two Higgs doublets, is chosen to be $5$, since larger values of $\tan\beta$ would give a problematically too short proton lifetimes (see Eqs.~\eqref{eq:CLLi}-\eqref{eq:CRL2})\footnote{In Ref.~\cite{Evans:2021hyx}, it is shown that the preferred value for minimal SUSY $SU(5)$ models is $\tan \beta \simeq 5$.}. 
The two independent GUT phases in the Yukawa couplings have a significant effect on the predicted proton lifetime. 
We, thus, scan over the GUT phases to find the maximum proton lifetime. 

\begin{figure}[!t]
\centering
\includegraphics[bb=0 0 450 457, width = 0.95\linewidth]{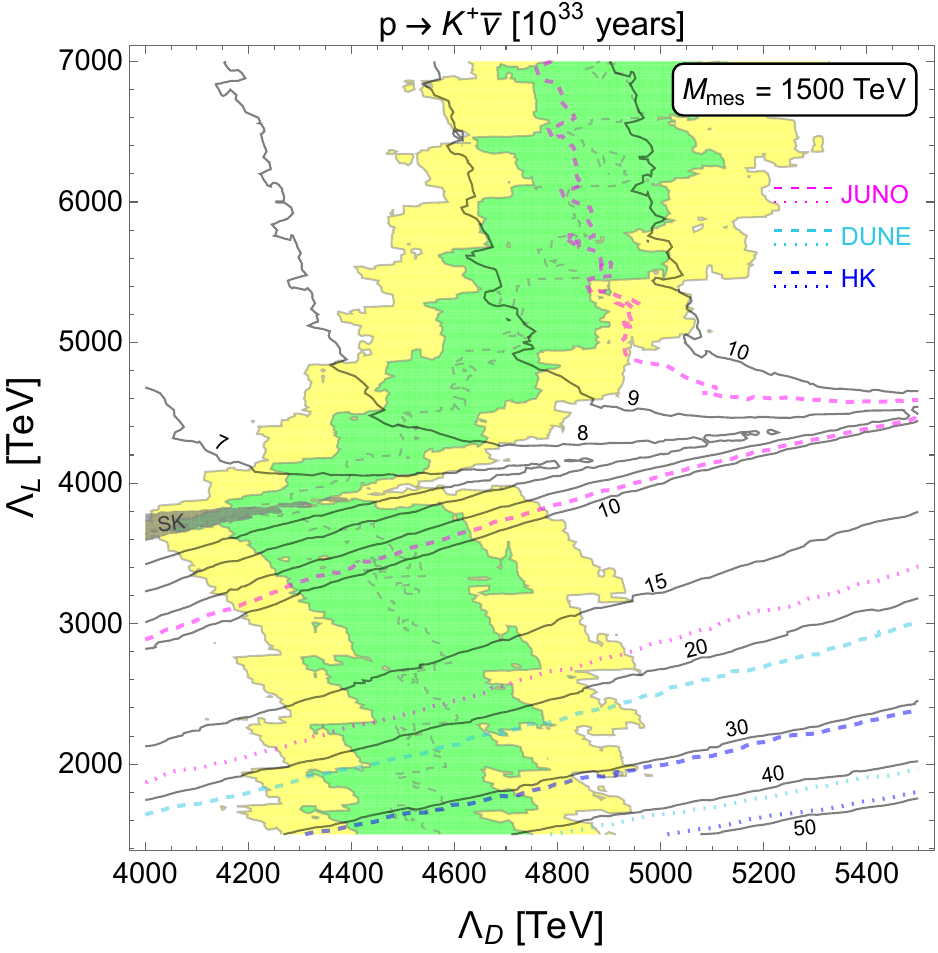}
\caption{Numerical result for proton lifetime of $\ptoKv$ process. 
The black lines correspond to $\tau (\ptoKv)$ in unit of $10^{33}$ years. 
The gray shaded area is excluded by the Super-Kamiokande experiment~\cite{Super-Kamiokande:2014otb}. 
Each experimental prospect for 10 (20) years runtime is also shown as dashed (dotted) lines, taken from Ref.~\cite{Bhattiprolu:2022xhm}. 
Green and yellow bands are $1\sigma$ and $2\sigma$ regions of observed Higgs mass, $125.20 \pm 0.11$ GeV~\cite{ParticleDataGroup:2024cfk}, and gray dashed line correspond to the central value of Higgs mass, $m_h = 125.20$ GeV.}
\label{fig:p2Kv}
\end{figure}

In Fig.~\ref{fig:p2Kv}, we show the predicted proton lifetime as well as future experimental prospects. 
Black contours correspond to the predicted lifetime for $\ptoKv$ in unit of $10^{33}$ years, and the gray shaded area is ruled out by the current experimental bound of $\ptoKv$, $5.9 \times 10^{33}$ years (at $90\%$ confidence level)~\cite{Super-Kamiokande:2014otb}. 
The blue, cyan and magenta contours are future prospects for 10 (dashed) and 20 (dotted) years runtime at Hyper-Kamiokande~\cite{Hyper-Kamiokande:2018ofw}, DUNE~\cite{DUNE:2015lol}, and JUNO~\cite{JUNO:2015zny}, respectively. 
Thus, most of the parameter space shown in Fig.~\ref{fig:p2Kv} can be tested at future experiments. 
However, to reproduce the SM Higgs mass, $125.20 \pm 0.11$ GeV~\cite{ATLAS:2015yey,CMS:2020xrn,ATLAS:2023oaq,ParticleDataGroup:2024cfk}\footnote{See the latest average reported in Particle Data Group 2024, \href{https://pdg.lbl.gov/2024}{https://pdg.lbl.gov/2024}.}, $\Lambda_D$ must fall within the green ($1\sigma$) and yellow ($2\sigma$) shaded regions roughly corresponding to $4000 \, {\rm TeV} \lesssim \Lambda_D \lesssim 5200 \, {\rm TeV}$, depending on the value of $\Lambda_L$. 
This constraint on $\Lambda_D$ is predominantly driven by the stop mass which is of order $\mathcal{O}(10)$ TeV in this range. 

It is clear that green and yellow bands for the Higgs mass have a non-trivial dependence on $\Lambda_{D, L}$. 
The key factor being the stop mass, which can be understood as follows. 
First, for the $\Lambda_L \lesssim 3200$ TeV region, the stop mass is determined solely by $\Lambda_D$, and hence, the Higgs mass bands are almost independent of $\Lambda_L$. 
For $3200 ~ {\rm TeV} \lesssim \Lambda_L \lesssim 4000$ TeV, the $\Lambda_L$ contribution to the stop masses and slightly smaller $\Lambda_D$ reproduce the needed stop masses. 
For $4000 ~ {\rm TeV} \lesssim \Lambda_D$, radiative corrections from stop mixing parameter ($X_t = A_t - \mu / \tan \beta$ with $A_t \equiv A_t (M_{\rm SUSY})$\footnote{Non-zero $A_t$ at the SUSY scale can be obtained by the RGE evolution from $M_{\rm mess}$ to $M_{\rm SUSY}$, even when we set all $A$-terms at $M_{\rm mess}$ to be zero.}) affect the prediction of the Higgs mass. 
This feature is mainly coming from the value of $\mu$: for $\Lambda_L \simeq 4000$ TeV, $\mu = \mathcal{O}(1)$ TeV, and increases with $\Lambda_L$. 
$A_t$, on the other hand, is almost completely determined by $\Lambda_D$. 
Hence, $X_t$ increases with $\Lambda_L$ for fixed $\Lambda_D$ in the range of $\Lambda_L > 4000$ TeV. 
We numerically check that the radiative corrections associated with $X_t$ decreases the SM Higgs mass for larger values of $\Lambda_L$. 
As a result, slightly larger $\Lambda_D$ is required to reproduce 125 GeV Higgs mass. 

Now, we discuss the dependence of the proton lifetime on $\Lambda_{D, L}$. 
From Fig.~\ref{fig:p2Kv}, it is seen that the proton lifetime is longest for larger $\Lambda_D$ and smaller $\Lambda_L$. 
The $\Lambda_D$ scaling is driven by the squark masses since they are proportional to $\Lambda_D$ and the amplitude $\ptoKv$ scales as $1 / M_{\rm SUSY}^2$ (see Eqs.~\eqref{eq:CLLi}-\eqref{eq:CRL2}). 
On the other hand, $\Lambda_L$ is related to the Wino mass which appears in the numerator of $C_{LL_i}$ as seen in Eq.~\eqref{eq:CLLi}. 
Thus, the lighter Wino mass is realized for smaller $\Lambda_L$ leading to a smaller proton decay amplitude and longer proton lifetime. 
However, $\Lambda_L$ contributes not only to the Wino mass but also left-handed squark masses. 
Thus, for fixed $\Lambda_D$, if the values of $\Lambda_L$ is large enough, the squarks begin to scale with $\Lambda_L$ resulting in a longer proton lifetime. 
For example, if $\Lambda_D = 4500$ TeV, we obtain the shortest lifetime, $\tau (\ptoKv) \simeq 6 \times 10^{33}$ years, for $\Lambda_L = 3950$ TeV and proton lifetime begins to grow for $\Lambda_L > 3950$ TeV. 

Importantly in the above result, we include contributions from the dimension-5 operator in Eq.~\eqref{eq:Weff}. 
If we set $d = 0$, we obtain a shorter proton lifetime due to the resulting relatively light colored Higgs mass, as studied in Ref.~\cite{Evans:2021hyx}. 
This case is excluded by the current limits. 
To justify the inclusion of the $\epsilon$ contribution, we have verified that it is less than $10^{-2}$ in entire parameter space shown in Fig.~\ref{fig:p2Kv}. 
This ensure that this operator does not require any large coefficient in order for us to find a viable model. 

Our final comment is on the prediction for the muon anomalous magnetic moment, $(g-2)_{\mu}$~\cite{Muong-2:2006rrc,Muong-2:2021ojo,Muong-2:2023cdq,Keshavarzi:2018mgv,Colangelo:2018mtw,Hoferichter:2019mqg,Davier:2019can,Keshavarzi:2019abf,Aoyama:2020ynm,Muong-2:2021ojo,Muong-2:2023cdq}. 
Since the minimal GMSB scenario we considered predicted relativity heavy Wino, Higgsinos and sleptons, all contributions to $(g-2)_{\mu}$ are too small. 
In the parameter space in Fig.~\ref{fig:p2Kv}, we obtain a $(g-2)_{\mu}$ contribution of $\mathcal{O}(10^{-12})$, which was calculated using \texttt{FeynHiggs 2.18.1}. 
Therefore, explaining the deviation in $(g-2)_{\mu}$ will require some kind of extensions to the minimal GMSB scenario. 
In future work, we will consider combining this with an explanation to the $(g-2)_{\mu}$ deviation.

\section{Summary}
\label{sec:summary}

In this paper, we discussed the proton lifetime in the minimal SUSY $SU(5)$ GUT model with GMSB scenario. 
We focused on the case of $\mathbf{5} + \mathbf{\bar{5}}$ messengers and calculate the mass spectrum for a messenger scale of $\mathcal{O}(10^3)$ TeV. 
Importantly, we found that UV completing these model with a GUT offered a solution to the $\mu$-$B_\mu$ problem which typically plagues gauge mediated models of supersymmetry breaking. 
This solution arises due to the fact that the dominant contribution to $B_\mu$ is from gravity mediation. 
With this being the case, we can use the same techniques that are used in gravity mediated models to get viable $\mu$ and $B_\mu$. 

In our study, we found parameter space consistent with both the SM Higgs mass measurements and current proton lifetime bounds. 
Also, we found that the predicted proton lifetime in the entire parameter space considered could be tested by future experiments such as Hyper-Kamiokande, DUNE and JUNO. 

In the minimal GMSB scenario, the desired value for $\Lambda_D$ is around 4500 TeV for $\Lambda_L \lesssim 4000$ TeV, which is necessary for reproducing the correct SM Higgs mass when the $A$-terms are small. 
Since smaller $\Lambda_L$ predicts a longer lifetime for $\ptoKv$, explicit breaking of the GUT relation between $\Lambda_L$ and $\Lambda_D$ is needed to obtain the experimentally measured Higgs mass. 
However, this can be realized by inclusion of a contribution to the messenger masses proportional to the GUT breaking field, $\Sigma$.

\vspace{0.5cm}




\appendix

\section{Details of the proton decay calculation}
\label{sec:calcinfo}

In this appendix, we summarize the relevant information for the calculation of proton decay. 
The other details can be found in, e.g., Refs.~\cite{Ellis:2015rya}. 
In our calculation, the relevant mode of proton decay is $\ptoKv_i$, whose scattering amplitude is given by
\begin{align}
&\mathcal{A} (\ptoKv_i) = \label{eq:Ap2Kvi} \\
&\hspace{3.5em} C_{LL_i} \left( \langle K^+ | (u s)_L d_L | p \rangle + \langle K^+ | (u d)_L s_L | p \rangle \right) \nonumber \\
&\hspace{2.2em}+ C_{RL_1} \langle K^+ | (u s)_R d_L | p \rangle + C_{RL_2} \langle K^+ | (u d)_R s_L | p \rangle \, , \nonumber
\end{align}
an approximate form for the coefficients $C_{LL_i}$ and $C_{RL_{1, 2}}$ can be found in Ref.~\cite{Ellis:2019fwf} are
\begin{align}
C_{LL_i} &\simeq \frac{2 \alpha_2^2}{\sin 2 \beta} \frac{m_t m_{d_i} M_2}{m_W^2 M_{H_C} M_{\rm SUSY}^2} V_{ui}^* V_{td} V_{ts} e^{i \phi_3} \nonumber \\
&\hspace{2.0em} \times \left( 1 + e^{i (\phi_2 - \phi_3)} \frac{m_c V_{cd} V_{cs}}{m_t V_{td} V_{ts}} \right) \, , \label{eq:CLLi} \\[0.5ex]
C_{RL_1} &\simeq - \frac{\alpha_2^2}{\sin^2 2 \beta} \frac{m_t^2 m_s m_{\tau} \mu}{m_W^4 M_{H_C} M_{\rm SUSY}^2} V_{tb}^* V_{us} V_{td} e^{i \phi_1} \, , \label{eq:CRL1} \\[0.5ex]
C_{RL_2} &\simeq - \frac{\alpha_2^2}{\sin^2 2 \beta} \frac{m_t^2 m_d m_{\tau} \mu}{m_W^4 M_{H_C} M_{\rm SUSY}^2} V_{tb}^* V_{ud} V_{ts} e^{i \phi_1} \, , \label{eq:CRL2}
\end{align}
where $\alpha_2 = g_2^2 / (4 \pi)$, $\tan \beta$ is the ratio of VEVs of two Higgs doublets, and $m_W$ is the $W$ boson mass. 
From recent results of the QCD lattice simulation~\cite{Aoki:2017puj}, relevant hadronic matrix elements are
\begin{align}
\langle K^+ | (u s)_L d_L | p \rangle &= 0.041(2)(5) \, {\rm GeV^2} \, , \label{eq:p2KusLdL} \\
\langle K^+ | (u d)_L s_L | p \rangle &= 0.139(4)(15) \, {\rm GeV^2} \, , \label{eq:p2KudLsL} \\
\langle K^+ | (u s)_R d_L | p \rangle &= -0.049(2)(5) \, {\rm GeV^2} \, , \label{eq:p2KusRdL} \\
\langle K^+ | (u d)_R s_L | p \rangle &= -0.134(4)(14) \, {\rm GeV^2} \, . \label{eq:p2KudRsL}
\end{align}

\bibliography{main}

\end{document}